\documentclass[%
 reprint,
superscriptaddress,
 amsmath,amssymb,
 aps,
]{revtex4-1}
\usepackage{xcolor}
\usepackage{graphicx}
\usepackage{dcolumn}
\usepackage{bm}
\usepackage{SIunits}
\usepackage{hyperref}

\begin{document}

\preprint{APS/123-QED}

\title{Tunable magnon-magnon coupling in synthetic antiferromagnets}

\author{A. Sud}
 \email{aakanksha.sud.17@ucl.ac.uk}
 \affiliation{
 London Centre for Nanotechnology, University College London, London WC1H 0AH, United Kingdom}
\author{C. W. Zollitsch}
 \affiliation{
 London Centre for Nanotechnology, University College London, London WC1H 0AH, United Kingdom}
\author{A. Kamimaki}
 \affiliation{Department of Applied Physics, Tohoku University, Aoba 6-6-05, Sendai, 980-8579, Japan}
 \affiliation{WPI Advanced Institude for Materials Research, Tohoku University, 2-1-1, Katahira, Sendai 980-8577, Japan}
 \author{T. Dion}
 \affiliation{
 London Centre for Nanotechnology, University College London, London WC1H 0AH, United Kingdom}
\author{S. Khan}
 \affiliation{
 London Centre for Nanotechnology, University College London, London WC1H 0AH, United Kingdom}
\author{S. Iihama}
 \affiliation{Frontier Research Institute for Interdisciplinary Sciences, Tohoku University, Sendai 980-8578, Japan}
  \affiliation{Center for Spintronics Research Network, Tohoku University, Sendai, 980-8577, Japan}
\author{S. Mizukami}
 \affiliation{WPI Advanced Institude for Materials Research, Tohoku University, 2-1-1, Katahira, Sendai 980-8577, Japan}
 \affiliation{Center for Spintronics Research Network, Tohoku University, Sendai, 980-8577, Japan}
 \affiliation{Center for Science and Innovation in Spintronics, Tohoku University, Sendai, 980-8577, Japan}
\author{H. Kurebayashi}%
 \email{h.kurebayashi@ucl.ac.uk}
 \affiliation{
 London Centre for Nanotechnology, University College London, London WC1H 0AH, United Kingdom}

\begin{abstract}
In this work, we study magnon-magnon coupling in synthetic antiferromagnets (SyAFs) using microwave spectroscopy at room temperature. Two distinct spin-wave modes are clearly observed and are hybridised at degeneracy points. We provide a phenomenological model that captures the coupling phenomena and experimentally demonstrate that the coupling strength is controlled by the out-of-plane tilt angle as well as the interlayer exchange field. We numerically show that a spin-current mediated damping in SyAFs plays a role in influencing the coupling strength.  
\end{abstract}

\maketitle


Generating new spin-wave states can be an enabling role for developing future spintronic/magnonic devices \cite{ChumacNaturePhys}. While individual spin-wave modes can be tailored by changing material parameters of host magnets, a novel approach of creating new spin-wave states is to couple two modes coherently by tuning them into resonance, where physical parameters of the coupled modes can also be modified. Although the coupling phenomena could be phenomenologically explained by a classical coupled-oscillator picture in general, microscopic descriptions of this type of hybridisation are rich, offering novel functionalities of state control and energy/information transfer. For example, strong coupling of light-matter interaction is envisaged to offer fast and protected quantum information processing \cite{KockumNatRevPhys2019,XiangRevModPhys2013,ClerkNatPhys2020}. Within this expanding research domain, strong coupling between microwave photons and collective spins in magnetically-ordered systems has been extensively studied in recent years \cite{Lachance-QuirionAPEX,HarderReview2018,HueblPRL2013,ZhangPRL2014}. 

Magnon-magnon coupling has an advantage over the light-matter interaction, in terms of coupling strength. The coupling strength of light-matter interactions is sometimes significantly reduced by a lack of spatial mode overlapping of the two, and so scientists have made considerable efforts to achieve large coupling strength by designing optimum geometries for efficient mode-volume overlapping \cite{EichlerPRL2017,TosiAIP2014}. On the other hand, magnon-magnon interaction does not suffer from this since two modes normally reside within the same host media, providing mode overlapping of 100\% or close to. While magnon-magnon coupling has been studied in single magnets \cite{Kalinikos1986,GrunbergBook,liensberger2019exchange,macneill2019gigahertz} and magnetic bi-layers \cite{ChenPRL2018,KlinglerPRL2018,LiPRL01}, magnon-magnon interaction in highly tunable material systems could offer unexplored parameter spaces on which to tailor the coupling phenomena. Here, we focus on synthetic antiferromagnets (SyAFs) as a host of magnon-magnon coupling and report clear hybridisation of two distinct SyAF modes arising from interlayer exchange coupling between two magnetic layers. We provide a full phenomenological model for the mode coupling, magnetic relaxation and coupling strength as a function of different material parameters for SyAF modes. Aided by these derived relationships, we demonstrate that the interlayer exchange field strength, which can be controlled by sample growth, allows the engineering of the coupling strength. We further numerically show that the spin-current mediated damping plays a role in influencing the coupling strength. Our demonstration and full details of the magnon-magnon coupling phenomena in SyAFs will act as a springboard for further research along this avenue \cite{DuineNPhys2018}.
\begin{figure}
\centering
\includegraphics[width=0.48\textwidth]{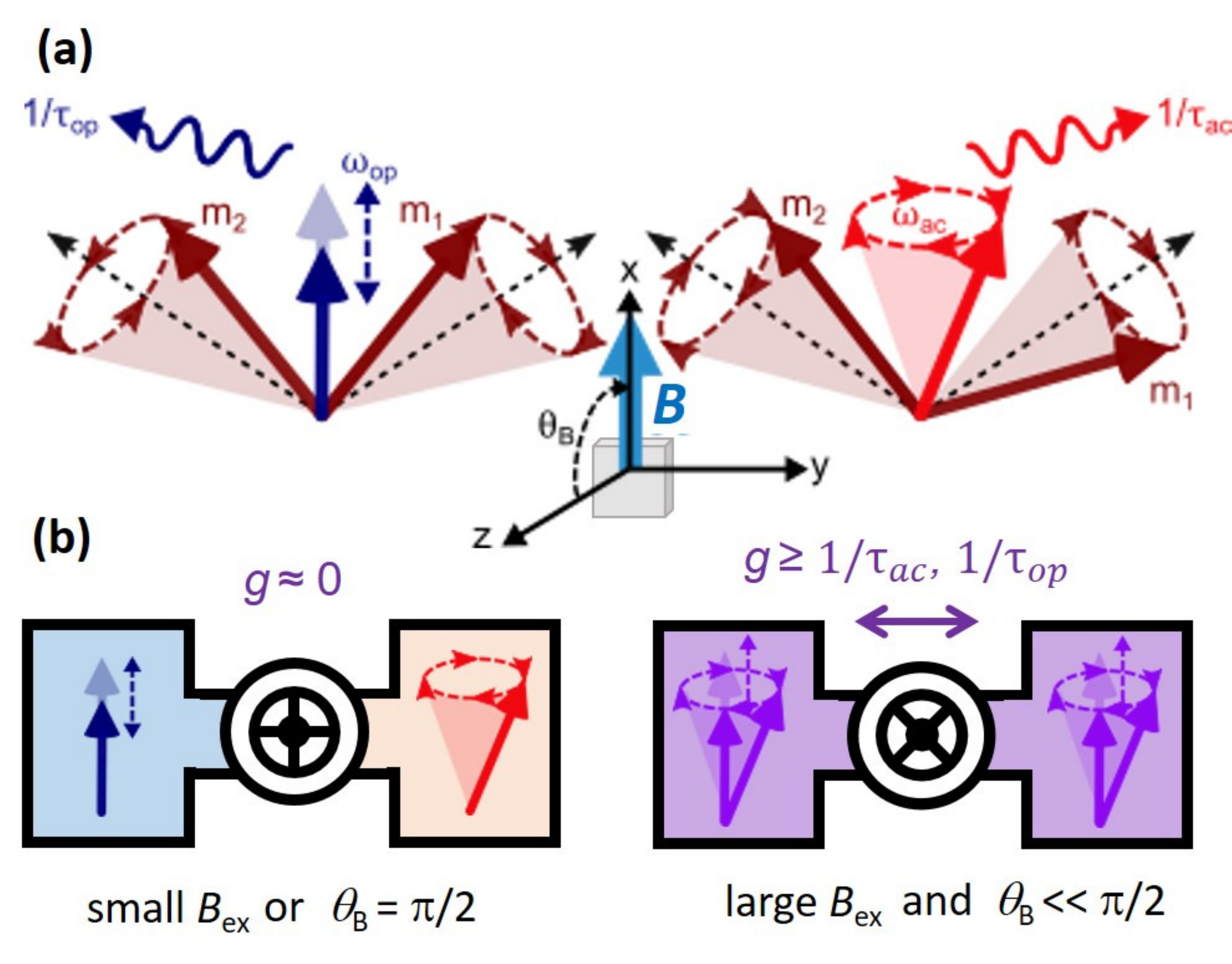}
\caption{\label{fig:fig_0}(a) Schematic of acoustic and optical modes in SyAFs. Two moments ($m_1$ and $m_2$) are coupled antiferromagnetically and canted at equilibrium. Under microwave irradiation, they precess in-phase (acoustic mode) and out-of-phase (optical mode) at different angular frequencies $\omega_{\text{ac}}$ and $\omega_{\text{op}}$, respectively. We define $\theta_\text{B}$ as in the figure, where the $z$ axis is the film growth direction. (b) Schematics of the magnon-magnon coupling phenomena with the optical and acoustic modes. When the exchange field ($B_\text{ex}$) is small or two moments are within the film plane, the coupling strength ($g$) is zero, so the two modes do not couple. We can valve the coupling strength by tuning $B_\text{ex}$ and $\theta_\text{B}$ and achieve strong magnon-magnon hybridisation, as shown on the right panel. }
\end{figure}

\begin{figure}
\centering
\includegraphics[width=0.48\textwidth]{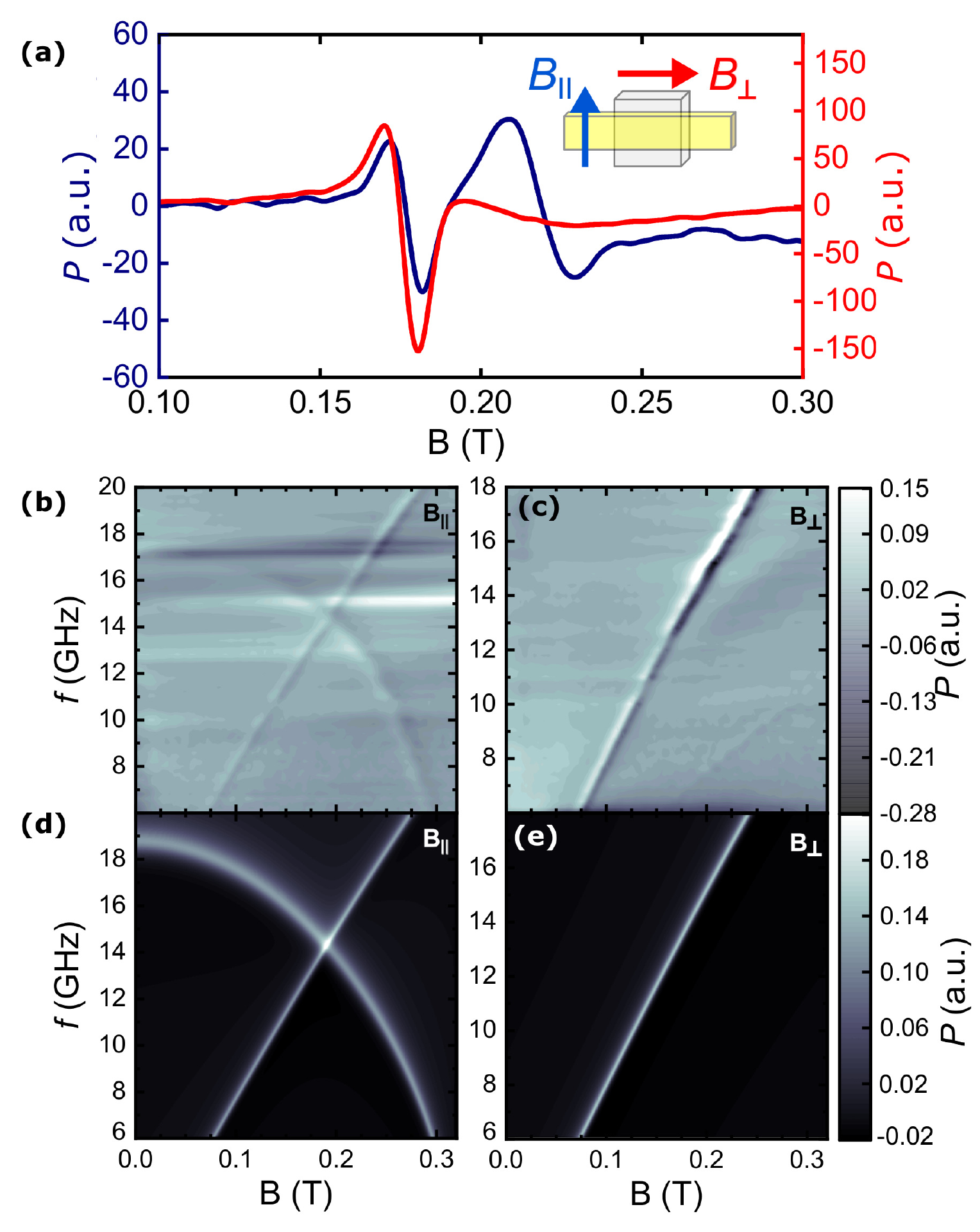}
\caption{\label{fig:fig_1}(a) Microwave absorption spectrum for $\theta_\text{B} = 90 ^{\circ}$, measured at 13.4 GHz. Two magnetic field directions ($B_\parallel$ and $B_\perp$) are defined as per the inset. Microwave transmission spectrum as a function of frequency and applied field for two configurations of applied magnetic fields (b) $B_\parallel$ and (c) $B_\perp$ for $\theta_\text{B} = 90 ^{\circ}$. (d)-(e) Theoretical results for the same experimental conditions as Fig. (b) and (c) respectively.}
\end{figure}

Low-energy spin-wave modes in synthetic antiferromagnets in their canted regime are acoustic and optical modes \cite{keffer1952theory,Rezende_JAP2019,Krebs1990} where two coupled moments precess in-phase (acoustic) and out-of-phase (optical) as shown in Fig.~\ref{fig:fig_0}(a). These modes have been studied and discussed already around 1990s, e.g.~by Grunberg et al. using Brilloin light scattering \cite{GrunbergPRL1984} and Zhang et al.~by microwave cavity experiments \cite{ZhangPRB1994}. The acoustic (optical) mode is excited by perpendicular (parallel) configuration between microwave and applied magnetic fields. There are a number of reports in which these two modes in different SyAFs have been studied in great detail \cite{KonovalenkoPRB2009,SekiAPL2009,Kamimaki_APL2019,WangAPL2018,IshibashiSciAdv2020}. For example, mutual spin pumping within the coupled moments has been proposed \cite{taniguchi2007enhancement,TakahashiAPL2014,chiba2015magnetization} and experimentally demonstrated \cite{TimopheevPRB2014,TanakaAPEX2014,YangAPL2016,SorokinPRB2020}. Both optical and acoustic mode frequencies as a function of the middle layer that influences the interlayer exchange coupling strength have been studied and reported earlier \cite{LiuPRB2014,SorokinPRB2020}. Since the resonant frequency of two modes shows different magnetic field dependence (as discussed more later), we can find the degeneracy point of the two modes by tuning experimental conditions. When the two moments are canted within the plane, the motion of the optical and acoustic modes can be decoupled \cite{macneill2019gigahertz}, meaning that the two modes are not allowed to hybridise. This restriction can be lifted when we tilt the moments towards the out-of-plane direction and we will be able to hybridise them (see Fig.1(b) for schematic understanding). The strength of hybridisation is defined by $g$ which represents a rate of energy transfer between the two modes. When this rate is fast, compared to mode dissipation rates of individual modes, we expect well-defined coupled modes before the excited states are relaxed. Control of the coupling strength $in$-$situ$ and $ex$-$situ$ will be potentially useful to a scheme of reconfigurable energy and information transfer using coherent coupling.  

\begin{figure*}[htp]
  \centering
  {\includegraphics[width=1.0\textwidth]{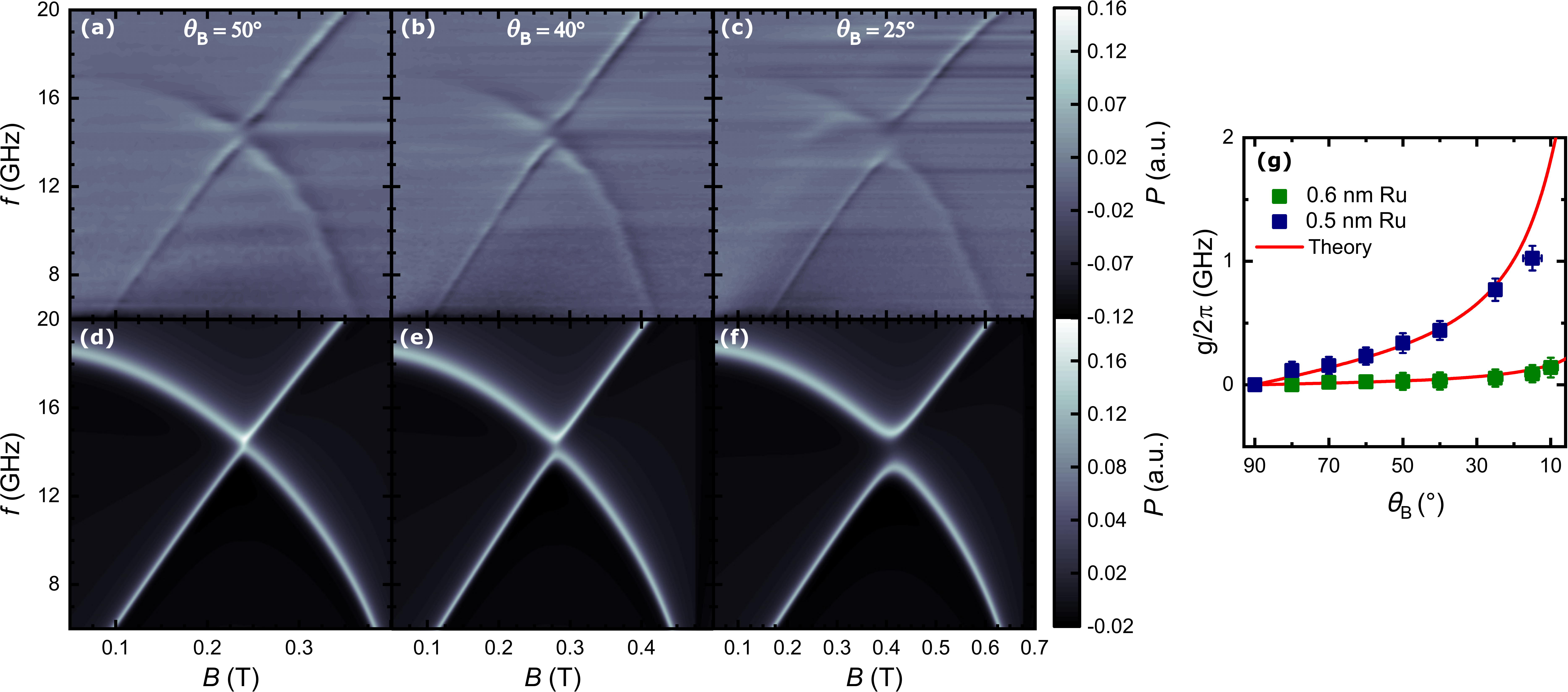}}\quad
  \caption{\label{fig:fig_2}(a-c) Microwave transmission as a function of frequency and applied field for different $\theta_\text{B}$. The avoided crossing starts to appear and the frequency gap increases as $\theta_\text{B}$ is decreased. (d)-(f) Simulation results for the same experimental condition as (a)-(c), respectively. (g) The coupling strength $g/2\pi$ as a function of $\theta_\text{B}$. We plot results from two samples with the Ru thickness of 0.5 nm and 0.6 nm. The 0.5 nm sample shows sizable $g/2\pi$, compared to much smaller $g/2\pi$ for 0.6 nm. The red curves are produced by Eq. (3) in the main text.}
\end{figure*}

The SyAF stacked films used in this study were prepared by magnetron co-sputtering at a base pressure of 1$\times$10$\textsuperscript{-7}$ Pa. The films were grown on a Si oxide substrate with the stacking pattern of Ta(3 nm)/CoFeB(3 nm)/Ru(t nm)/CoFeB(3 nm)/Ta(3 nm) where Ru thickness was varied to tune the interlayer exchange coupling \cite{Kamimaki_APL2019}. Vibrating sample magnetometer (VSM) was used to characterise the static magnetic properties (see Supplemental Material\cite{supplemental}). These sample chips were placed on a coplanar waveguide board to perform broadband spin dynamics characterisation. For each measurement, we fixed the frequency and swept a dc external magnetic field with an ac modulation component at 12 Hz. Figure~\ref{fig:fig_1}(a)  shows typical measurement curves for two field directions ($B_\parallel$ and $B_\perp$) defined by the figure inset. We carried out systematic experiments for a wide range of frequency (5-20 GHz) as well as field angle to study mode hybridisation and linewidth evolution of SyAFs. To extract the peak position and linewidth, we used derivative Lorentzian functions \cite{Rogdakis_PRM2019}. Figures~\ref{fig:fig_1} (b)-(c)  show two-dimensional color plots of microwave absorption as a function of microwave frequency and magnetic field. We can clearly identify two modes in Fig.~\ref{fig:fig_1}(b)  for the $B_\parallel$ condition whereas only one in Fig.~\ref{fig:fig_1}(c)  for $B_\perp$. This is because for $B_\parallel$, both modes can be excited since microwave rf fields have components of both parallel and perpendicular to $B_\parallel$ due to their spatial distribution above the waveguide. For $B_\perp$ measurements, the microwave magnetic field only possesses components perpendicular to $B_\perp$, hence only exciting the acoustic mode. In order to analyse these results quantitatively, we solve a coupled Landau-Lifshtitz-Gilbert (LLG) equation with small angle approximation \cite{streit1980antiferromagnetic,vs1973antiferromagnetic,macneill2019gigahertz,KamimakiPRAppl2020} (see Supplemental Material\cite{supplemental} for more details) and find the resonance condition of the two modes as:
\begin{equation}
\omega_{ac} = \gamma B_0\sqrt{\left(1+\frac{B_{\text{s}}}{2B_{\text{ex}}}\right)},\label{eq:two}\\
\end{equation}
\begin{equation}
 \omega_{op} = \gamma \sqrt{2B_{\text{ex}}B_{\text{s}}\left(1-\left(\frac {B_0}{2B_{\text{ex}}}\right)^{2}\right)},
 \label{eq:three}
\end{equation}

Here, $B_{\text{ex}}$, $B_{\text{s}}$, $B_{0}$ and $\gamma$ are the exchange field, the demagnetisation magnetization, the resonance field and the gyromagnetic ratio, respectively. We found that our best fits produce $B_{\text{ex}}$, $B_{\text{s}}$ and $\gamma/2 \pi$ to be 0.14 T, 1.5 T and 29 GHz/T respectively. Resonance frequencies predicted by  Eqs.~\ref{eq:two} and \ref{eq:three} can reproduce our experimental results very well as shown in Figs. \ref{fig:fig_1}(d) and (e), strongly supporting that we can experimentally observe and study the coupled SyAF modes. Since the frequency of the two modes show different magnetic field dependences, it is possible to study mode coupling of the two by tuning the mode frequencies. In Fig.~\ref{fig:fig_1} (b) , we observe a clear crossing of the two modes at $B_0 \approx 0.2$ T. This "crossing" means that the two modes are not able to hybridise due to mode symmetry \cite{macneill2019gigahertz}. We can break this symmetry by tilting the moment towards the out-of-plane direction. We therefore repeated similar experiments for $\theta_\text{B} \neq 90^{\circ}$ as shown in Figs.~\ref{fig:fig_2}(a-c). The two modes start to show an avoided crossing as $\theta_\text{B}$ is decreased, indicating mode hybridisation which can be quantitatively discussed by using the coupling strength $g/2\pi$, the half of the minimum frequency gap. We plot the $\theta_\text{B}$ dependence of $g/2\pi$ in Fig.~\ref{fig:fig_2}(g) where $g/2\pi$ grows with the out-of-plane component, with the highest value exceeding 1 GHz.

We describe the magnon-magnon coupling phenomena in SyAFs by a 2$\times$2 matrix eigenvalue problem derived from the coupled LLG equations with mutual spin pumping terms \cite{chiba2015magnetization} (see Supplemental Material\cite{supplemental} for detailed derivation):
\[ \begin{array}{c}
\left[ \begin{array}{cc}
{\mathrm{\omega }}^{\mathrm{2}}-{\mathrm{\omega }}^{\mathrm{2}}_\text{op}+i(\nu_\text{o1}+\nu_\text{o2})\mathrm{\omega } & \left(i\mathrm{\omega }-{\nu }_\text{o1}\gamma B_s\right)\eta m_\text{z0} \\ 
\left(-i\mathrm{\omega }+{\nu }_\text{a2}\gamma B_s\right)\eta m_\text{z0} & {\mathrm{\omega }}^{\mathrm{2}}-{\mathrm{\omega }}^{\mathrm{2}}_\text{ac}+i(\nu_\text{a1}+\nu_\text{a2})\mathrm{\omega } \end{array}
\right]
\end{array}
\]
Here, $\eta=2B_{\text{ex}}/B_\text{s}$, $m_\text{z0}=B_0 \text{cos} \theta _\text{B}/(B_\text{s}+2B_\text{ex})$, ${\nu }_\text{o1}=(\alpha_0+\alpha_\text{sp})(1-m_\text{z0}^2)-\alpha_\text{sp}\{1-m_\text{z0}^2-(B_0^2\text{sin}^2 \theta _\text{B}/4B_\text{ex}^2)\}(m_\text{z0}^2/m^2)$, ${\nu }_{o2}=\alpha_0 \eta(1-B_0^2\text{sin}^2 \theta _\text{B}/4B_\text{ex}^2)$, ${\nu }_\text{a1}=\alpha_0 \eta(m_{z0}^2+B_0^2\text{sin}^2 \theta _\text{B}/4B_\text{ex}^2)$ and ${\nu }_{a2}=\alpha_0(\eta+1)(1-m_{z0}^2)$, respectively, with $\alpha_0$ and $\alpha_\text{sp}$ being the standard Gilbert damping constant and one arising from mutual spin pumping between the two magnetic layers. The real part of the eigenvalues gives the resonance frequencies and the imaginary part represents the loss rates of the two modes. We numerically solved the eigenvalue problem with parameters described above and found that the coupled equations can model our experimental observation well for each experimental set, such as Figs. ~\ref{fig:fig_2} (d)-(f) reproducing corresponding experimental results. We simplified the 2$\times$2 matrix by neglecting the damping terms to calculate the eigenvalues and found an analytical expression for the coupling strength as (see derivation in Supplemental Material\cite{supplemental}):

\begin{equation}\label{eq:coupling}
g\mathrm{=}\frac{\gamma B_{\text{ex}}B_0}{2B_\text{s}+4B_{\text{ex}}}{\mathrm{cos} {\theta }_\text{B}}. 
\end{equation}

This correctly captures our experimental observation as $g/2\pi$ grows with decreasing $\theta_\text{B}$. The red curve in Fig. \ref{fig:fig_2}(g) is calculated by this equation and there is quantitative agreement between experiments and theory, despite marginal deviation at small $\theta_\text{B}$. To further attest the validity of this equation for our experiments, we performed similar measurements on a SyAF sample having the Ru thickness of 0.6 nm since Eq. \ref{eq:coupling} suggests that the coupling strength can be tuned by $B_{\text{ex}}$. For this sample, we found that $B_{\text{ex}}$ is decreased to 30 mT due to a weaker interlayer coupling and accordingly, as expected, we observed a significant decrease of $g/2\pi$ as summarised in Fig. \ref{fig:fig_2}(g). These results show the tunability of the mode coupling strength in SyAFs by both thin-film growth engineering (\textit{ex-situ}) as well as out-of-plane tilt angle (\textit{in-situ}).

\begin{figure*}[htp]
  \centering
  {\includegraphics[width= 1.0\textwidth]{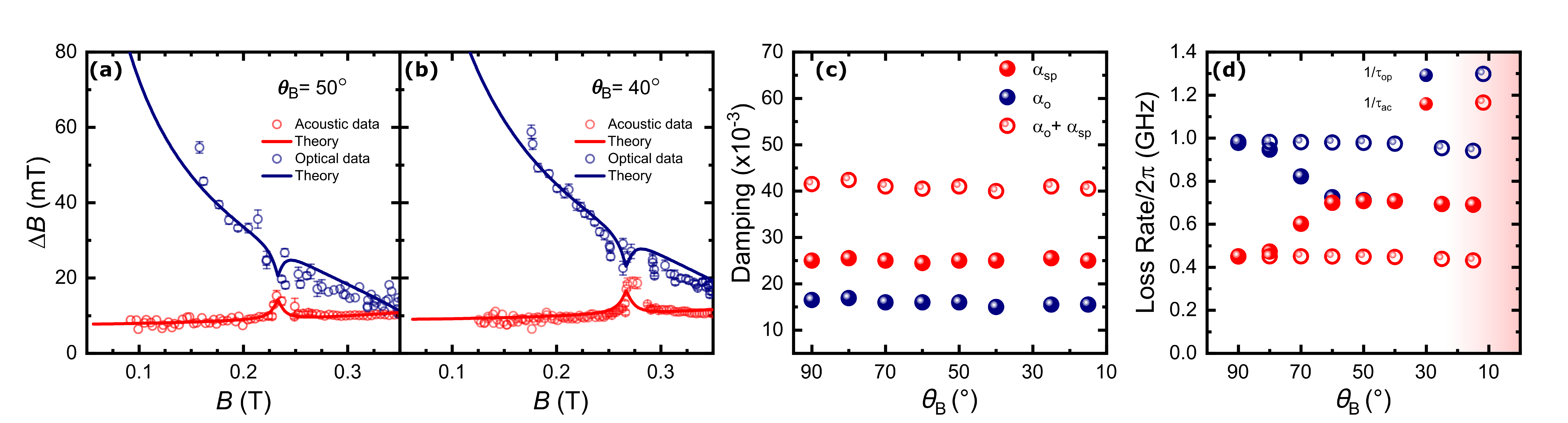}}\quad
  \caption{\label{fig:fig_3}(a-c) HWHM linewidth as a function of magnetic field for $\theta_B$ of (a)50$^\circ$ and (b) 40$^\circ$. solid lines represent results from the theoretical model discussed in the main text. (c) Extracted values of damping parameters. (d) Calculated loss rates of each mode at the crossing point as well as those of the hybridised modes.
 }
\end{figure*}

Next we focus on the relaxation of the SyAF modes. Figure~\ref{fig:fig_3}(a-b) represent plots of the half width at half maximum (HWHM) linewidth ($\Delta B$) extracted for individual sweeps for both modes. $\Delta B$ of the acoustic mode increases with increasing magnetic field, with a characteristic anormaly around the field where the two modes hybridise. $\Delta B$ of the optical mode however shows a different magnetic field dependence as it decreases with increasing magnetic field. This is primarily due to the relationship of the magnetic-field-domain linewidth and frequency-domain linewidth as given by:
\begin{equation}\label{eq:linewidth}
\mathrm{\Delta }B_{\mathrm{op(ac)}}={\left|\frac{d{\omega }_{\mathrm{op(ac)}}}{dB}\right|}^{-1}\frac{1}{{\tau }_{\mathrm{op(ac)}}}.
\end{equation}
When the resonance field is low, $\lvert d{\omega }_{\mathrm{op}}/dB \rvert$ becomes small, which can extrinsically enhance the observed $\Delta B$ in our experiments. In order to extract material-specific parameters such as $\alpha_0$ from our data, we solved the eigenvalue problem and compared the imaginary part with experimental results. We found that the linewidth calculated from the imaginary part models excellently for our experiments as shown in Figs. 4(a)-(b). Extracted $\alpha_0$ and $\alpha_{_\text{sp}}$ for different $\theta_\text{B}$ are plotted in Fig. \ref{fig:fig_3} (c). We can confirm that there is a sizable spin pumping component for every angle we measured, in consistent with previous reports \cite{chiba2015magnetization,YangAPL2016,Kamimaki_APL2019,HeinrichPRL2003}. The Ru thickness is much shorter than its spin diffusion length of 14 nm \cite{Eid_JAP2002}. As a result, when two ferromagnets are precessing in-phase, according to spin pumping theory \cite{tserkovnyak2002enhanced}, spin currents flowing out of the two are cancelled out hence developing zero time-dependent spin accumulation in the Ru layer. However, when two moments precess out of phase, the emitted spin currents no longer cancel out, leading to the spin-accumulation which induces an additional damping mechanism for the optical mode. In our experiments, we observe that both $\alpha_0$ and $\alpha_{_\text{sp}}$ are independent of $\theta_\text{B}$, which can be understood that the Gilbert damping components are a material parameter, independent of experimental conditions - note here that the canted angle has been already taken into account in the expressions.

An interesting observation is that the experimentally deduced $\Delta B$ for both modes also show "attraction" around the avoided crossing points. This demonstrates that magnetic relaxation can be modified by mode coupling phenomena. In the crossing regime, two modes are no longer pure acoustic or optical and therefore it is not possible to use the ac spin pumping picture associated with the phase difference between two moments. Rather, a simple phenomenological picture of hybridised energy losses would be a better one. When two modes with different loss rates start to couple coherently, their loss rates also start to merge together \cite{CarmichaelPRA1989}. This is because the energy transfer mixes the two loss rates since the high (low)-loss mode becomes the low(high)-loss mode as a function of time. We are able to observe this feature in our experiments. This loss rate hybridisation is reproduced by our numerical simulations from the eigenvalue problem as shown in Figs.~\ref{fig:fig_3}(a-b). This linewidth averaging is similar to ones discussed in spin-photon coupling systems \cite{HarderPRB2017,HarderPRL2018} as well as magnon-magnon coupling at YIG/NiFe interfaces \cite{LiPRL01}.  We went on to quantify the loss rates for both modes by using Eq. \ref{eq:linewidth}. First of all, we estimated the loss rate of individual modes at the avoided crossing point (open circles in Fig. 4 (d)), by extrapolating from the values outside the coupling regime. Both show a very weak angular dependence, which can be understood that the damping (Fig.4 (c)) has no angular dependence with a subtle change of the mode-crossing frequency when $\theta_\text{B}$ is decreased. By contrast, loss rates for the hybridised modes (solid circles in Fig. 4 (d)), estimated by our eigenvalue problem, exhibit clear attraction as the coupling strength is increased by changing $\theta_\text{B}$. After $\theta_\text{B} = 60^{\circ}$, the loss rates of the two modes coalesce into a single number which is  exactly the average of the two rates 1/$\tau_{\text{mix}}$=(1/2)(1/$\tau_{\text{ac}}$+1/$\tau_{\text{op}}$) where 1/$\tau_{\text{mix}}$ is the loss rate of the hybridised states. Furthermore, through the course of our simulation study, we found that $\alpha_{_\text{sp}}$ can have an effect on $g$, suggesting that the magnon-magnon coupling is partially mediated by spin currents. We observe that for large $\alpha_{_\text{sp}}$ the coupling between the two modes can be completely suppressed (see Supplemental Material\cite{supplemental}). We highlight that this damping-mediated coupling control cannot be achieved by simply changing $\alpha_0$ in our system, something specific for the magnetic relaxation via spin pumping to the coupling and the energy exchange. Although it is not possible to control $\alpha_{_\text{sp}}$ in our experiments, it could act as an extra parameter to define the magnon-magnon coupling strength in SyAFs. Finally, we highlight that the highest $g/2\pi$ achieved (1.0 GHz) outnumbers the loss rates of the individual modes, indicating that this magnon-magnon coupling starts to enter the strong coupling regime in our experiments. Although our experiments are just at the onset of the strong coupling regime, here we briefly discuss potential improvements and control of the coupling strength against the individual loss rates. Equation (3) can be simplified as $g/2\pi \propto B_\text{ex}/B_{\text{s}}$, suggesting that a sample with a higher $B_\text{ex}$ as well as a smaller $B_\text{s}$ shows a large coupling strength. Achieving similar coupling with low-damping materials could be another plausible path. 

In summary, we experimentally show the magnon-magnon coupling in SyAF CoFeB/Ru/CoFeB multi-layers. Clear magnon-magnon hybridisation has been observed when the optical and acoustic modes are tuned into resonance. The magnon-magnon coupling strength has been controlled by bringing the moments into the out-of-plane direction, which breaks the orthogonality of the two modes. In addition, the interlayer exchange coupling is found to tune the coupling strength. The loss rate of two modes exhibits an averaging effect upon hybridisation. Our eigenvalue problem approach serves to provide the analytical expression of the coupling strength as well as numerical explanations/predictions of the experimental data. We envisage that results in the present study will be transferable to other weakly-coupled antiferromagnetic systems since the phenomenological descriptions of their spin-wave modes should be identical to our model developed.
$\textit{Note added}$. We became aware that recently similar magnon-magnon coupling in synthetic antiferromagnets has been observed for finite wavelength spin-waves by Shiota et al. \cite{shiota2020tunable}. 

A.K. acknowledges the Graduate Program in Spintronics (GP-Spin) at Tohoku University. This work was supported in part by CSRN, CSIS and UCL-Tohoku Strategic Partner Funds.



\providecommand{\noopsort}[1]{}\providecommand{\singleletter}[1]{#1}%
%
%

\pagebreak
\widetext
\begin{center}
\textbf{\large Supplementary Material for "Tunable magnon-magnon coupling in synthetic antiferromagnets"}

\end{center}
\renewcommand{\theequation}{S\arabic{equation}}
\renewcommand{\thefigure}{S\arabic{figure}}
\setcounter{figure}{0}

\makeatletter
\renewcommand{\theequation}{S\arabic{equation}}
\renewcommand{\thefigure}{S\arabic{figure}}
\renewcommand{\bibnumfmt}[1]{[S#1]}
\renewcommand{\citenumfont}[1]{S#1}

\def\ve{\varepsilon}
\def\unit#1{\ \mathrm{#1}}
\font\myfont=cmr12 at 16pt

\begin{figure}[b]
\centering
{\includegraphics[width=0.7\textwidth]{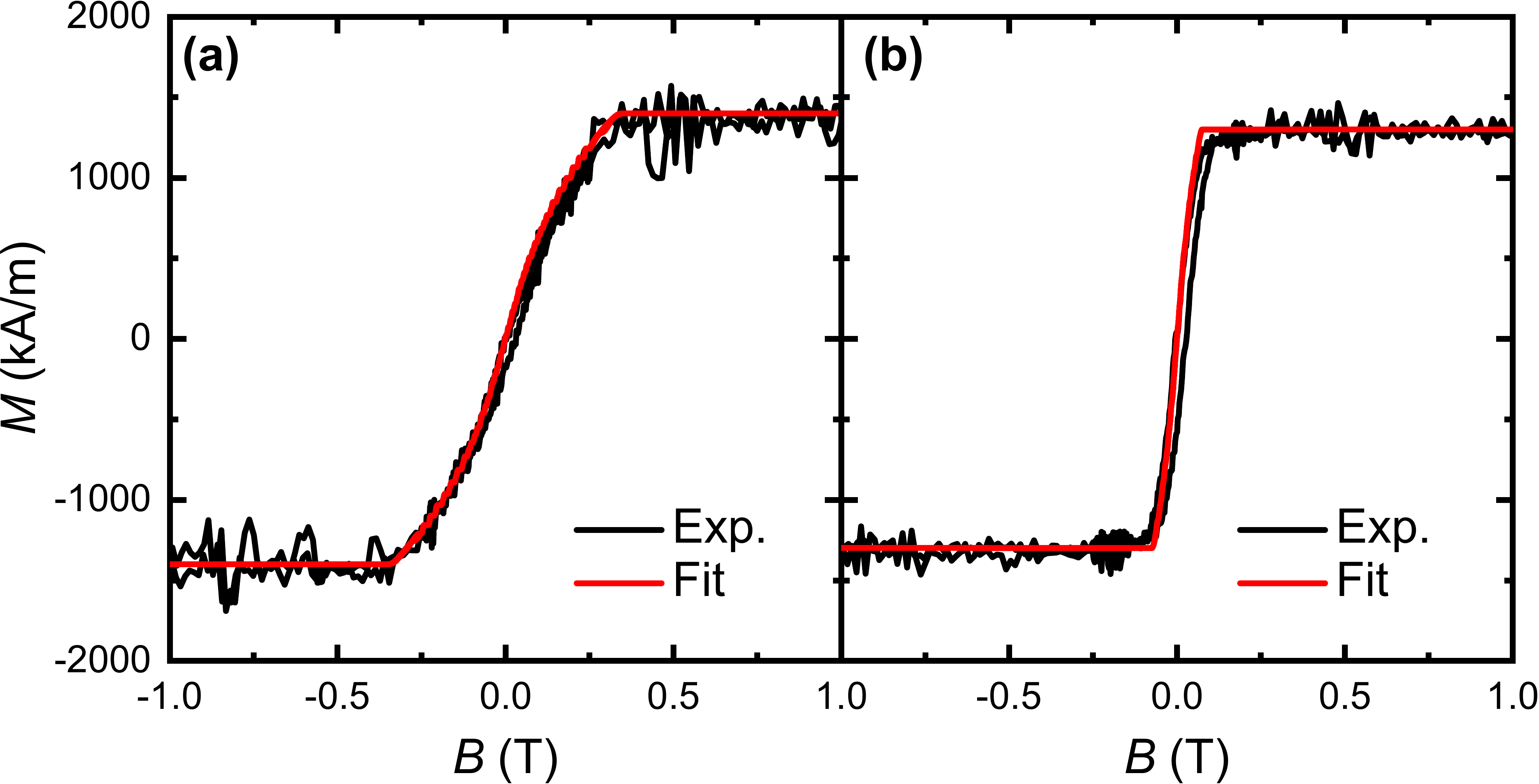}}
\caption{\label{fig:fig_s1}(a-b) Magnetization curve of the CoFeB (3 nm)/Ru ($t$ nm)/CoFeB (3 nm) measured by vibrating sample magnetometer for (a) $t$ = 0.5 and (b) $t$ = 0.6. The black (red) curve is the experimental (calculation) results.}
\end{figure}
\section{Sample characterization by vibrating sample magnetometer}

In this study, we use the following free energy expression which includes of linear and quadratic exchange coupling contributions \cite{biquadratic,martin,sorokin} to describe static magntisation direction in a synthetic antiferromagnet (SyAF):
 \[ \begin{array}{c}\tag{S1}
 F=\sum\limits_{j=1}^{2}\bigg[M_\text{s}\boldsymbol{B}\cdot\boldsymbol{m_j}+\frac{1}{2}M_\text{s}B_\text{s}\left(\boldsymbol{m_j}\boldsymbol{\cdot }\boldsymbol{z}\right)^2\bigg]+\frac{2J_\text{ex1}}{d}\boldsymbol{m_1}\cdot\boldsymbol{m_2}+\frac{2J_\text{ex2}}{d}\left(\boldsymbol{m_1}\cdot\boldsymbol{m_2}\right)^2
 . \end{array}
\] 

\noindent Here, $M_\text{s}$, ${\boldsymbol{B}}$, ${\boldsymbol{m_{1(2)}}}$, $B_\text{s}$, $J_\text{ex1(2)}$ are the saturation magnetisation, external magnetic field vector, the unit vector of individual moments in a SyAF, demagnetisation field, the linear and quadratic antiferromagnetic interlayer exchange coupling constants, respectively; $d$ is the thickness of two ferromagnetic layers which are identical in the present case. Figure~\ref{fig:fig_s1} shows magnetometery characterisation of two samples used in the present study. The red lines in the figure are calculated by using $M\left(B\right)=M_s\text{cos}\phi (B)$ \cite{sorokin,parkin,martin} where $\phi(B)$ is the angle between the applied magnetic field direction and equilibrium direction of individual moments which is obtained for all field values by minimizing Eq. S1 reiteratively until we achieve good matching to experimental data. The red curves in Fig. S1 were generated by the linear and quadratic exchange fields of 140 (30)  $\pm$ 1.2 (0.6) mT and 7 (2) $\pm$ 0.1 (0.03) mT for the 0.5 (0.6) nm Ru thickness sample, together with $M_s $= 1400 (1300) kA/m  for the 0.5 (0.6) nm Ru sample. The effective magnetic field acting on both moments can be given by differentiating the exchange coupling terms ($F_\text{ex}$) in Eq. S1 with respect to $\boldsymbol m_{1(2)} $:
\[ \begin{array}{c}\tag{S2}
 B_\text{ex,1(2)}=-\frac{1}{2M_\text{s}}\frac{\partial F_\text{ex}}{\partial m_{1(2)}}=-\frac{J_\text{ex1}}{d}\boldsymbol{m_{2(1)}}-\frac{2J_\text{ex2}}{d}\left(\boldsymbol{m_1}\cdot\boldsymbol{m_2}\right)\boldsymbol m_{2(1)}
 , \end{array}
\] 
$\boldsymbol m_1 \cdot \boldsymbol m_2 $ is a scalar value defined by the relative angle between $\boldsymbol m_1$ and $\boldsymbol m_2$. Note that we incorporate this second-order exchange coupling term within $B_\text{ex}$ for our analysis in our study. 

\section{Additional magnetisation-dynamics results in this study}

This section provides supplementary results used in our study to support our claims in the main text. In Fig. S2, we show a number of our individual scans for microwave absorption experiments. In Fig. S3, we show 2D plots of frequency vs magnetic field for different $\theta_\mathrm{B} $ from the sample with the Ru thickness of 0.5 nm. This supplements Fig. 3 in the main text and further supports our observation of crossing/avoided-crossing feature, 
\begin{figure}[h]
\centering
{\includegraphics[width=0.6\textwidth]{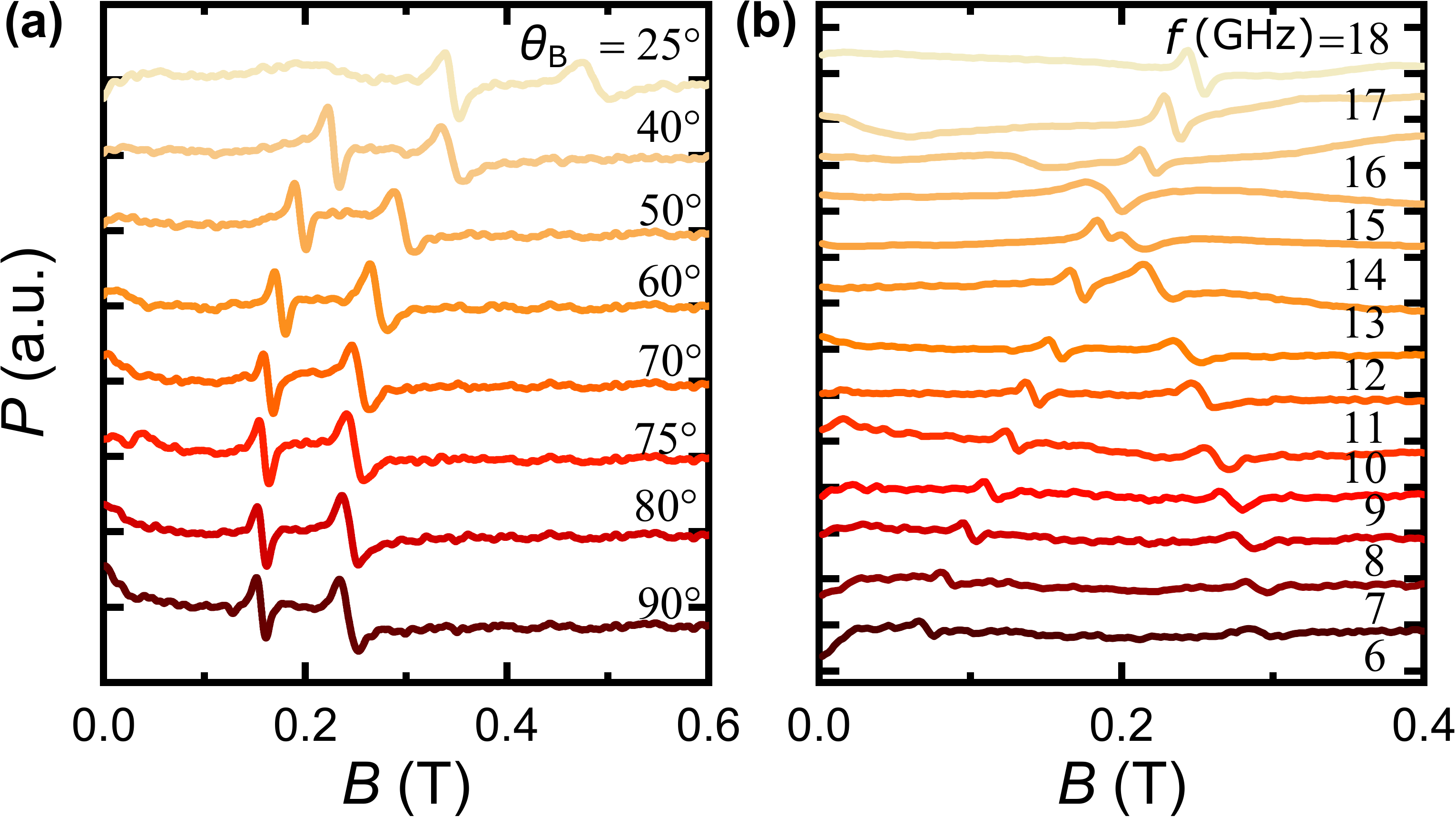}}\label{fig:sRu0.5_0}
\caption{\label{fig:fig_s2_0} Individual field scans taken by measuring our Ru 0.5 nm sample for (a) different $\theta_\mathrm{B}$ and excitation frequency of 12
GHz, and (b) for different frequencies and $\theta_\mathrm{B}$ = 90$^\degree$
.
 }
\end{figure}

\begin{figure}[h]
\centering
{\includegraphics[width=0.96\textwidth]{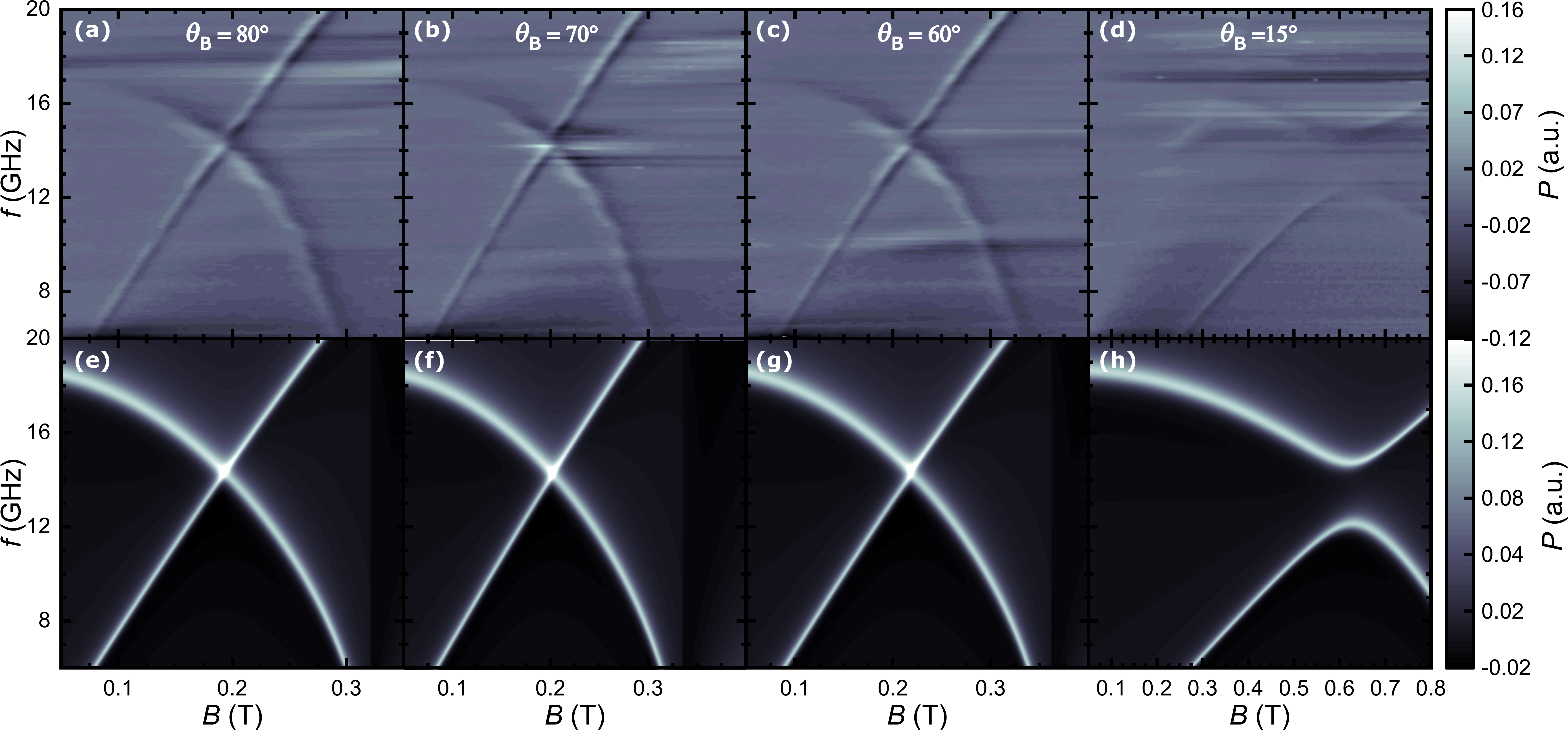}}\label{fig:sRu0.5}
\caption{\label{fig:fig_s2}(a-d)Extra data plots of Microwave transmission as a function of frequency and applied field, for the sample with the Ru thickness of 0.5 nm for different $\theta_\mathrm{B}$. Large coupling gap can be seen at low angles. Figures (e-h) plot simulation results for the same experimental conditions as Fig.(a-d). }
\end{figure}
\begin{figure}[h]
\centering
{\includegraphics[width=1\textwidth]{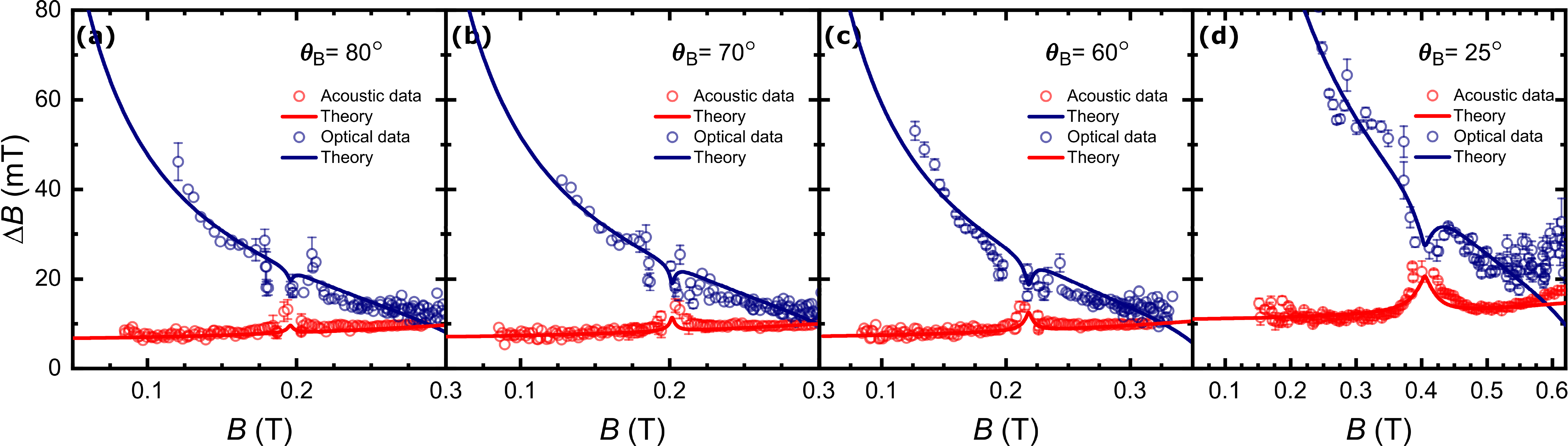}}
\caption{\label{fig:fig_s3}(a-d) Extra data plots of linewidth of the two modes as a function of magnetic field for different  $\theta_\mathrm{B}$ for the sample with the Ru thickness of 0.5 nm. Solid lines represent simulation results from the theoretical model we used in this study.}
\end{figure}
\newpage
\noindent controlled by the out-of-plane angle $\theta_\text{B}$ in the main text. Furthermore, Fig. S4 represents the magnetic-field-domain linewidth ($\Delta B$) as a function of frequency measured for different $\theta_\text{B}$. Theory curves plotted were produced by the imaginary part of eigenvalues discussed in the main text and Section 3 in this document. Damping parameters and loss rates plotted in Fig. 4(d) in the main text have been extracted from the parameters in the eigenvalue problem. 

We repeated similar measurements for the sample with the Ru thickness of 0.6 nm. The same analysis procedure and plots have been carried out for experimental data and shown in Figs. S5 and S6. In Fig. S5, we notice that there exist magnetic-field independent background signals around 5 GHz which we consider as transmission losses unrelated to magnetisation dynamics. Nevertheless, we here highlight that the gap opening is much weaker than those measured for the sample with the Ru thickness of 0.5 nm. This is attributed to the size of exchange coupling, which has been independently quantified by VSM as explained above.   

\begin{figure}[h]
\centering
{\includegraphics[width=1\textwidth]{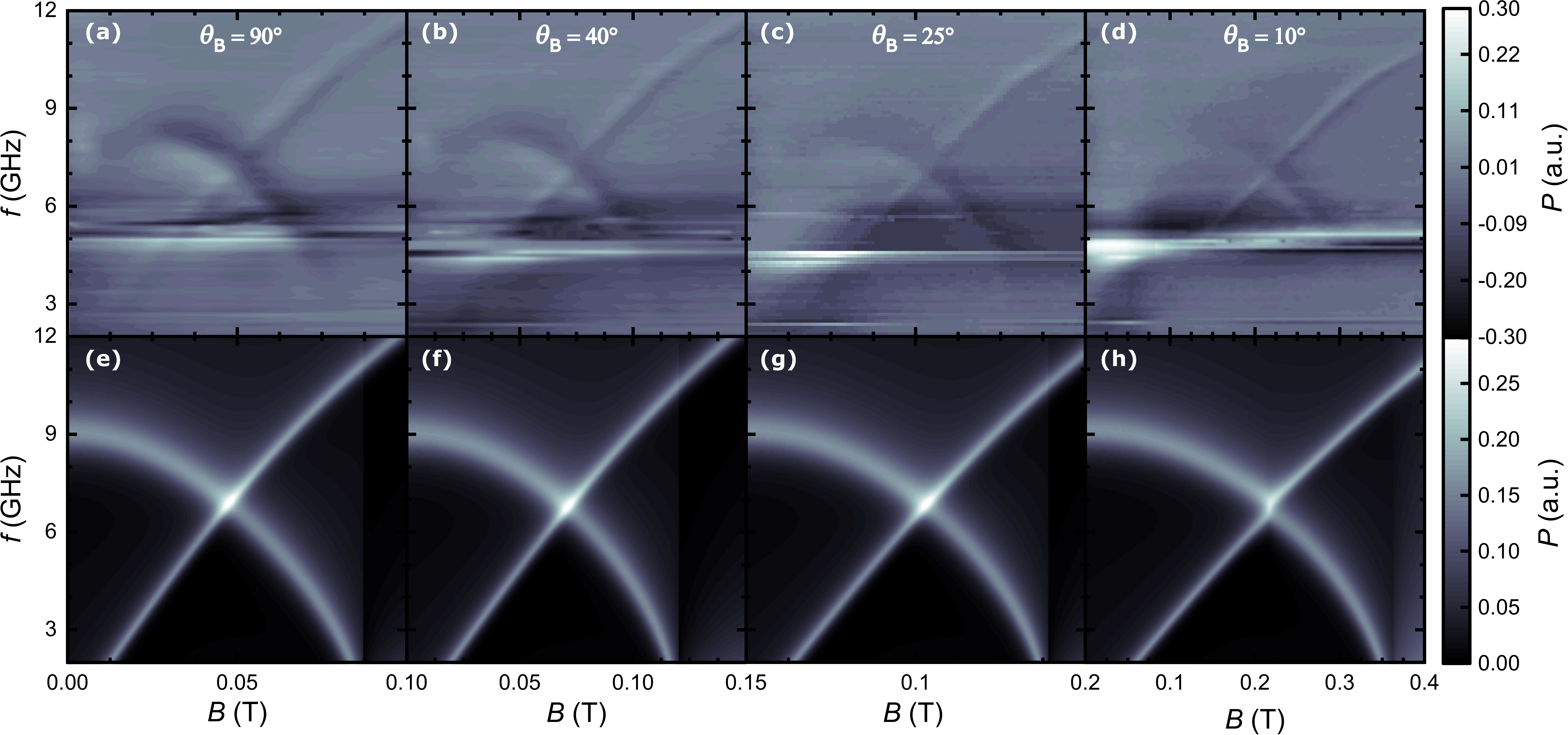}}
\caption{\label{fig:fig_s4}(a-d)Microwave transmission as a function of frequency and applied field, for the sample with the Ru thickness of 0.6 nm for different $\theta_\mathrm{B}$. Small gap opening corresponds to the weak exchange coupling of the sample. Figures (e-h) plot simulation results for the same experimental conditions as Fig. (a-d). }
\end{figure}
\begin{figure}[h!]
\centering
{\includegraphics[width=1.0\textwidth]{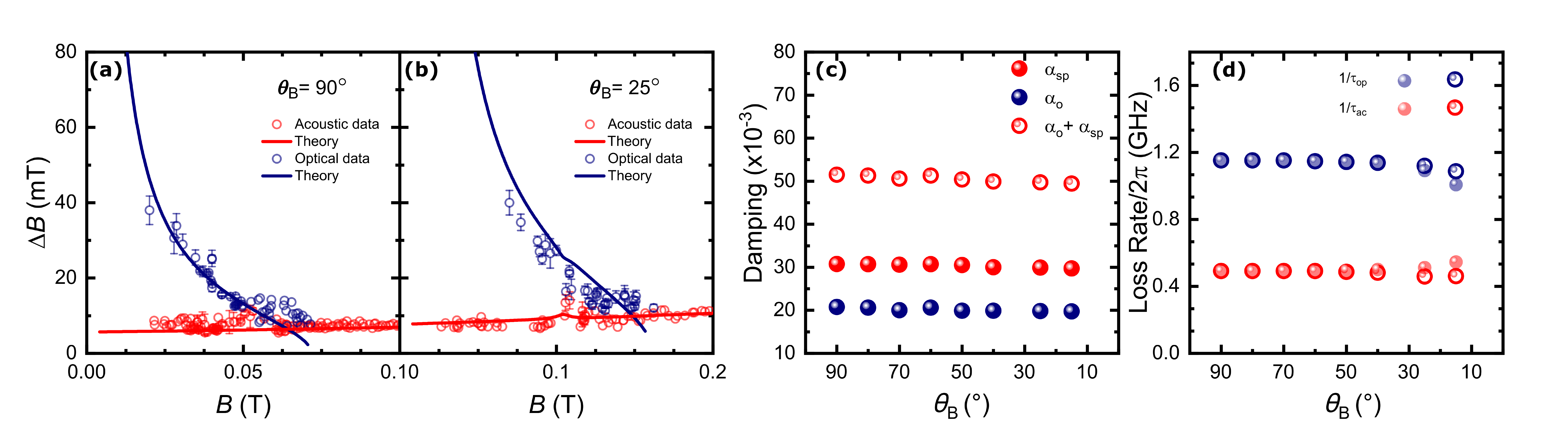}}
\caption{\label{fig:fig_s5}(a-b) Linewidth of the two modes as a function of magnetic field for $\theta_\mathrm{B}$ of (a) 90$^\degree$ and (b) 25$^\degree$ for the sample with the Ru thickness of 0.6 nm. We show our simulation results as solid lines. (c) Extracted values of damping parameters. (d) Calculated loss rates of each mode at the crossing point as well as those of the hybridised modes.}
\end{figure}

\newpage

\section{The eigenvalue problem and analytical expressions}
In this section, the two-coupled Landau-Lifshitz-Gilbert (LLG) equations at the macrospin limit are employed to model magnetic dynamics of optical and acoustic modes in SyAFs. We recoginse that similar approaches have been taken by others previously \cite{Kamimaki_APL2019,chiba2015magnetization,macneill2019gigahertz} but not specifically for magnon-magnon coupling phenomena in SyAFs as we detail below. We consider a canted regime of two individual moments ($\boldsymbol m_1$ and $\boldsymbol m_2$) which are coupled antiferromagnetically by the exchange interaction with the strength of $B_\text{ex}$. These two moments reside in thin-film magnets subjected to a demagnetisation field $B_\text{s}$ and we apply an external magnetic field ${\boldsymbol{B}}$ within the $x$-$z$ plane with angle $\theta_\text{B}$ from the $z$ axis which is the sample growth direction in our case. Following convention, we first define Kittel and Neel vectors as $\boldsymbol m= \left( \boldsymbol m_1 +\boldsymbol m_2\right)/2$ and $\boldsymbol n= \left( \boldsymbol m_1 -\boldsymbol m_2\right)/2$, respectively. Dynamics of these two coupled moments are given by \cite{chiba2015magnetization}:
\[ \begin{array}{c}\tag{S3}
\frac{d\boldsymbol{m}}{dt}=-{\mathrm{\Omega }}_\text{L}\boldsymbol{m}\times\boldsymbol{u}\boldsymbol{+}{\mathrm{\Omega }}_\text{B}\left[\left(\boldsymbol{m}\boldsymbol{\cdot }\boldsymbol{z}\right)\boldsymbol{m}\times\boldsymbol{z}+\left(\boldsymbol{n}\boldsymbol{\cdot }\boldsymbol{z}\right)\boldsymbol{n}\times\boldsymbol{z}\right]\boldsymbol{+}{\boldsymbol{\tau}}_\text{m}, \end{array}
\] 
\[ \begin{array}{c}\tag{S4}
\frac{d\boldsymbol{n}}{dt}=-{\mathrm{\Omega }}_\text{L}\boldsymbol{n}\times\boldsymbol{u}\boldsymbol{+}{\mathrm{\Omega }}_\text{B}\left[\left(\boldsymbol{m}\boldsymbol{\cdot }\boldsymbol{z}\right)\boldsymbol{n}\times\boldsymbol{z}+\left(\boldsymbol{n}\boldsymbol{\cdot }\boldsymbol{z}\right)\boldsymbol{m}\times\boldsymbol{z}\right]+2{\mathrm{\Omega }}_\text{ex}\boldsymbol{n}\times\boldsymbol{m}\boldsymbol{+}{\boldsymbol{\tau}}_\text{n}.\ \end{array}
\] 
Here, $ {\mathrm{\Omega }}_\text{L}=\gamma B_0$, ${\mathrm{\Omega }}_\text{B}=\gamma B_\text{s}$, and ${\mathrm{\Omega }}_\text{ex}=\gamma B_\text{ex}$ where $\gamma$ and $B_\text{0}$ are the gyromagnetic ratio and the resonance field, respectively; ${\boldsymbol{u}}= \text{sin}\theta_\text{B}{\boldsymbol{x}} + \text{cos}\theta_\text{B}{\boldsymbol{z}}$ represents the applied field direction in the $x$-$z$ plane, given that ${\boldsymbol{x}}$ and ${\boldsymbol{z}}$ are unit vectors for the corresponding axes. The last terms in Eqs. (S4) and (S5) account for damping torques which are expressed as:
\[ \begin{array}{c}\tag{S5}
{\boldsymbol{\tau}}_\mathrm{m}={\alpha }_0\left(\boldsymbol{m}\times\frac{d\boldsymbol{m}}{dt}+\boldsymbol{n}\times\frac{d\boldsymbol{n}}{dt}\right),\ \end{array}
\] 
\[ \begin{array}{c}\tag{S6}
{\boldsymbol{\tau}}_n=\left({\alpha }_\mathrm{0}+{\alpha }_\mathrm{sp}\right)\left(\boldsymbol{m}\times\frac{d\boldsymbol{n}}{dt}\boldsymbol{+}\boldsymbol{n}\times\frac{d\boldsymbol{m}}{dt}\right)-{\alpha }_\mathrm{sp}\left[\boldsymbol{m}\boldsymbol{\cdot }\left(\boldsymbol{n}\times\frac{d\boldsymbol{m}}{dt}\right)\frac{\boldsymbol{m}}{m^2~}\boldsymbol{+}\boldsymbol{n}\boldsymbol{\cdot }\left(\boldsymbol{m}\times\frac{d\boldsymbol{n}}{dt}\right)\frac{\boldsymbol{n}}{n^2~}\right],\end{array}
\] 

where $\alpha_0$ and $\alpha_\text{sp}$ are the standard Gilbert damping constant and one arising from mutual spin pumping between the two magnetic layers. With approximation of small angle precession, we can separate the equilibrium 
($\boldsymbol{m}_{\boldsymbol{0}}$ and $\boldsymbol{n}_{\boldsymbol{0}}$) and time-dependent ($\delta \boldsymbol{m}\left(t\right)$ and $\delta \boldsymbol{n}\left(t\right)$) terms as $\boldsymbol{m}\left(t\right)={\boldsymbol{m}}_{\boldsymbol{0}}\boldsymbol{+}\delta \boldsymbol{m}\left(t\right)$ and $\boldsymbol{n}\left(t\right)={\boldsymbol{n}}_{\boldsymbol{0}}\boldsymbol{+}\delta \boldsymbol{n}\left(t\right)$ and here we define each vector component in a standard manner, $e.g.$ ${\boldsymbol{m}}_{\boldsymbol{0} }= (m_\mathrm{0x},m_\mathrm{0y},m_\mathrm{0z})$. Substituting $\boldsymbol{m}\left(t\right)$ and $\boldsymbol{n}\left(t\right)$ into Eqs. (S4)-(S7) and keeping the first order of the time-dependent and damping terms, we obtain the following six coupled equations:

\[ \begin{array}{c}\tag{S7}
\frac{1}{{\mathrm{\Omega }}_\mathrm{B}}\frac{d\delta m_\mathrm{x}}{dt}=n_\mathrm{0y}\delta n_\mathrm{z}-{\alpha }_\mathrm{0}\eta \left(m^2_\mathrm{0z}+n^2_\mathrm{0y}\right)\delta m_\mathrm{x}-\eta m_\mathrm{0z}\delta m_\mathrm{y}+{\alpha }_\mathrm{0}(\eta+1) m_\mathrm{0x}m_\mathrm{0z}\delta m_\mathrm{z},\end{array}
\] 
\[ \begin{array}{c}\tag{S8}
\frac{1}{{\mathrm{\Omega }}_B}\frac{d\delta m_\mathrm{y}}{dt}={\alpha }_0m_\mathrm{0z}n_\mathrm{0y}\delta n_\mathrm{z}+\eta m_\mathrm{0z}{\delta m}_x-{\alpha }_0\eta m^2_0{\delta m}_y-(\eta+1) m_\mathrm{0x}\delta m_\mathrm{z}, \end{array}
\] 
\[ \begin{array}{c}\tag{S9}
\frac{1}{{\mathrm{\Omega }}_B}\frac{d\delta m_\mathrm{z}}{dt}={\alpha }_0\eta m_\mathrm{0x}m_\mathrm{0z}{\delta m}_x+\eta m_\mathrm{0x}{\delta m}_y-{\alpha }_0(\eta+1) \left(m^2_{0x}+n^2_{0y}\right)\delta m_\mathrm{z}.\end{array}
\] 
\[ \begin{array}{c}\tag{S10}
\frac{1}{{\mathrm{\Omega }}_B}\frac{d\delta n_\mathrm{x}}{dt}=(\alpha_0+\alpha_\text{sp})m_\mathrm{0x}m_\mathrm{0z}\delta n_\mathrm{z}+\alpha_0\eta n_\mathrm{0y}m_\mathrm{0x}{\delta m}_y+(\eta+1)n_\mathrm{0y}\delta m_\mathrm{z}+\alpha_\mathrm{sp}\frac{m_\mathrm{0z}m_\mathrm{0x}n^2_\mathrm{0y}}{m^2_\mathrm{0}}\delta n_\mathrm{z},\end{array}
\] 
\[ \begin{array}{c}\tag{S11}
\frac{1}{{\mathrm{\Omega }}_B}\frac{d\delta n_\mathrm{y}}{dt}=-m_\mathrm{0x}\delta n_\mathrm{z}+\alpha_0\eta m_\mathrm{0x}n_\mathrm{0y}\delta m_\mathrm{x}+\alpha_0(\eta+1) m_\mathrm{0z}n_\mathrm{0y}\delta m_\mathrm z, \end{array}
\] 
\[ \begin{array}{c}\tag{S12}
\frac{1}{{\mathrm{\Omega }}_B}\frac{d\delta n_\mathrm{z}}{dt}=~-(\alpha_0+\alpha_\text{sp})\left((m^2_{0x}+n^2_{0y}\right)\delta n_\mathrm{z}-\eta n_\mathrm{0y}\delta m_\mathrm{x}+\alpha_0\eta m_\mathrm{0z}n_\mathrm{0y}\delta m_\mathrm{y}+\alpha_\mathrm{sp}\frac{m^2_\mathrm{0z}n^2_\mathrm{0y}}{m^2_\mathrm{0}}\delta n_\mathrm{z}, \end{array}
\]

where the equilibrium conditions and a new parameter $\eta$ are introduced: ${\boldsymbol{m}}_{\boldsymbol{0} }= (m_\mathrm{0x},0,m_\mathrm{0z})$, ${\boldsymbol{n}}_{\boldsymbol{0} }= (0,n_\mathrm{0y},0)$ and $\eta =2{\mathrm{\Omega }}_{ex}/ {\mathrm{\Omega }}_B$. Further simplification has been made by using additional parameter definitions as well as time derivative treatment:
\[ \begin{array}{c}\tag{S13}
{\nu }_\mathrm{o1}=(\alpha_0+\alpha_\text{sp})\left(m^2_\mathrm{0x}+n^2_\mathrm{0y}\right)-\alpha_\mathrm{sp}\frac{m^2_\mathrm{0z}n^2_\mathrm{0y}}{m^2_\mathrm{0}}\end{array}
\] 
\[ \begin{array}{c}\tag{S14}
{\nu }_\mathrm{o2}={\alpha }_0\eta \left(m^2_\mathrm{0z}+n^2_\mathrm{0y}\right), \end{array}
\] 
\[ \begin{array}{c}\tag{S15}
{\nu }_\mathrm{a1}={\alpha }_0\eta m^2_0,
\end{array}
\] 
\[ \begin{array}{c}\tag{S16}
{\nu }_\mathrm{a2}={\alpha }_0(\eta+1) \left(m^2_\mathrm{0x}+n^2_\mathrm{0y}\right),
\end{array}
\]
\[\begin{array}{c}\tag{S17}
\frac{1}{{\mathrm{\Omega }}_\mathrm{B}}\frac{d}{dt}\to -i\frac{\omega }{{\mathrm{\Omega }}_\mathrm{B}}\equiv -i\mathrm{\Omega }.
\end{array}
\]

These produce the following equations of motion for $\delta m_\mathrm{x}$, $\delta m_\mathrm{y}$, $\delta m_\mathrm{z}$ and $\delta n_\mathrm{z}$:

\[ \begin{array}{c}\tag{S18}
-i\mathrm{\Omega }\delta m_\mathrm{x}=n_\mathrm{0y}\delta n_\mathrm{z}-{\nu }_\mathrm{o2}\delta m_\mathrm{x}-\eta m_\mathrm{0z}\delta m_\mathrm{y}+{\alpha }_0(\eta+1) m_\mathrm{0x}m_\mathrm{0z}\delta m_\mathrm{z},
\end{array}
\] 
\[ \begin{array}{c}\tag{S19}
-i\mathrm{\Omega }\delta m_\mathrm{y}={\alpha }_0m_\mathrm{0z}n_\mathrm{0y}\delta n_\mathrm{z}+\eta m_\mathrm{0z}{\delta m}_x-{\nu }_\mathrm{a1}{\delta m}_y-(\eta+1) m_\mathrm{0x}\delta m_\mathrm{z},
\end{array}
\] 
\[ \begin{array}{c}\tag{S20}
-i\mathrm{\Omega }\delta m_\mathrm{z}={\alpha }_0\eta m_\mathrm{0x}m_\mathrm{0z}{\delta m}_x+\eta m_\mathrm{0x}{\delta m}_y-{\nu }_\mathrm{a2}\delta m_\mathrm{z},
\end{array}
\] 
\[ \begin{array}{c}\tag{S21}
-i\mathrm{\Omega }\delta n_\mathrm{z}=~-{\nu }_\mathrm{o1}\delta n_\mathrm{z}-\eta n_\mathrm{0y}\delta m_\mathrm{x}+\alpha_0\eta m_\mathrm{0z}n_\mathrm{0y}\delta m_\mathrm{y}.
\end{array}
\] 
We write these as a matrix form as follows.
\[ \begin{array}{c}\tag{S22}
\mathrm{\Omega }\left( \begin{array}{c}
\delta n_\mathrm{z} \\ 
{\delta m}_\mathrm{x} \\ 
\delta m_\mathrm{y} \\ 
{\delta m}_\mathrm{z} \end{array}
\right)=i\left( \begin{array}{cccc}
-{\nu }_\mathrm{o1} & -\eta n_\mathrm{0y} & \alpha_0\eta m_\mathrm{0z}n_\mathrm{0y} & 0 \\ 
n_\mathrm{0y} & -{\nu }_\mathrm{o2} & -\eta m_\mathrm{0z} & {\alpha }_0(\eta+1) m_\mathrm{0x}m_\mathrm{0z} \\ 
{\alpha }_0m_\mathrm{0z}n_\mathrm{0y} & \eta m_\mathrm{0z} & -{\nu }_\mathrm{a1} & -(\eta+1) m_\mathrm{0x} \\ 
0 & {\alpha }_0\eta m_\mathrm{0x}m_\mathrm{0z} & \eta m_\mathrm{0x} & -{\nu }_\mathrm{a2} \end{array}
\right)\left( \begin{array}{c}
\delta n_\mathrm{z} \\ 
{\delta m}_\mathrm{x} \\ 
\delta m_\mathrm{y} \\ 
{\delta m}_\mathrm{z} \end{array}
\right)
\end{array}
\] 
We can obtain the eigen mode frequency $\omega_{op/ac}$ and relaxation time $\tau_{op/ac}$ by solving the complex eigenvalue problem, which can be expressed as,

\[ \begin{array}{c}\tag{S23}
\Omega_\mathrm{op(ac)}=\frac{1}{\Omega_\mathrm{B}}(\omega_\mathrm{op(ac)}-i{\tau}^{-1}_\mathrm{op(ac)}).
\end{array}
\] 

If we neglected damping terms in the off-diagonal components, we obtained the following relations:
\[ \begin{array}{c}\tag{S24}
\delta n_\mathrm{z}\sim -\frac{\eta n_\mathrm{0y}}{-i\mathrm{\Omega }+{\nu }_\mathrm{o1}}\delta m_\mathrm{x}.
\end{array}
\] 
\[ \begin{array}{c}\tag{S25}
\delta m_\mathrm{z}\sim \frac{\eta m_\mathrm{0x}}{-i\mathrm{\Omega }+{\nu }_\mathrm{a2}}{\delta m}_\mathrm{y},
\end{array}
\] 
Using these relations, we reduced Eq. (S23) into an eigenvalue problem with a 2-by-2 matrix form given by:
\[ \begin{array}{c}\tag{S26}
\left( \begin{array}{cc}
{\mathrm{\Omega }}^{\mathrm{2}}-{\mathrm{\Omega }}^{\mathrm{2}}_\mathrm{op}+i(\nu_\text{o1}+\nu_\text{o2})\mathrm{\Omega } & -\left(-i\mathrm{\Omega }+{\nu }_\mathrm{o1}\right)\eta m_\mathrm{0z} \\ 
\left(-i\mathrm{\Omega }+{\nu }_\mathrm{a2}\right)\eta m_\mathrm{0z} & {\mathrm{\Omega }}^{\mathrm{2}}-{\mathrm{\Omega }}^{\mathrm{2}}_{ac}+i(\nu_\text{a1}+\nu_\text{a2})\mathrm{\Omega } \end{array}
\right)\left( \begin{array}{c}
{\delta m}_\mathrm{x} \\ 
\delta m_\mathrm{y} \end{array}
\right)= 0
\end{array}
\] 
During this process, we disregarded higher-order terms in the diagonal elements, such as ${\nu }_\mathrm{o1}{\nu }_\mathrm{o2}$. This is the matrix we show in the main text after converting $\Omega$ into $\omega$ using Eq. S25. Here, ${\mathrm{\Omega }}_{\mathrm{op}}$ and ${\mathrm{\Omega }}_{\mathrm{ac}}$ are the eigen frequencies for optical mode and acoustic modes given by:
\[ \begin{array}{c}\tag{S27}
\mathrm{\Omega_\text{op} }=n_\mathrm{0y}\sqrt{\eta }-i\frac{1}{2}\left({\nu }_\mathrm{o1}+{\nu }_\mathrm{o2}\right),
\end{array}
\] 
\[ \begin{array}{c}\tag{S28}
\mathrm{\Omega_\text{ac} }=\sqrt{\eta^2+\eta}m_\mathrm{0x}-i\frac{1}{2}\left({\nu }_\mathrm{a1}+{\nu }_\mathrm{a2}\right).
\end{array}
\] 
  
Equations (1) and (2) in the main text can be obtained by the real part of these two equations by using the equilibrium conditions: $(m_\mathrm{0x},m_\mathrm{0y},m_\mathrm{0z})=(B_0 \text{sin} \theta_\text{B}/2B_\text{ex},0,B_0 \text{cos} \theta_\text{B}/(B_\text{s}+2B_\text{ex})) $ and $(n_\mathrm{0x},n_\mathrm{0y},n_\mathrm{0z})=(0,\sqrt{1-m^2_{0x}-m^2_{0z}},0) $ with $\theta_\text{B}=\pi/2$. We note here that our numerical solutions of (S23) and (S27) are almost identical and therefore we decided to show the simpler 2$\times$2 matrix in the main text. In order to provide the coupling constant $g$ in the main text, we further take out the damping terms and solved the following eigen problem.
\[ \begin{array}{c}\tag{S29}
\left( \begin{array}{cc}
{\mathrm{\Omega }}^{\mathrm{2}}-{\mathrm{\Omega }}^{\mathrm{2}}_\mathrm{op} & i\mathrm{\Omega }\eta m_\mathrm{0z} \\ 
-i\mathrm{\Omega }\eta m_\mathrm{0z} & {\mathrm{\Omega }}^{\mathrm{2}}-{\mathrm{\Omega }}^{\mathrm{2}}_\mathrm{ac} \end{array}
\right)\left( \begin{array}{c}
{\delta m}_\mathrm{x} \\ 
\delta m_\mathrm{y} \end{array}
\right)
\end{array}=0
\] 
\[ \begin{array}{c}\tag{S30}
\left({\mathrm{\Omega }}^2-{\mathrm{\Omega }}^2_\mathrm{op}\right)\left({\mathrm{\Omega }}^2-{\mathrm{\Omega }}^2_\mathrm{ac}\right)-{\mathrm{\Omega }}^2{\left(\eta m_\mathrm{0z}\right)}^2=0
\end{array}
\] 
By defining the crossing (dimensionless) frequency as ${\mathrm{\Omega }}_{\mathrm{op}}={\mathrm{\Omega }}_{\mathrm{ac}}={\mathrm{\Omega }}_0$, we can find the energy gap ($\mathrm{\Delta }{\mathrm{\Omega}_{\mathrm{gap}}}$) using Eq. (S32). 
\[ \begin{array}{c}\tag{S31}
\mathrm{\Delta }{\mathrm{\Omega }}_{\mathrm{gap}}\mathrm{=}2(\mathrm{\Omega }\mathrm{-}{\mathrm{\Omega }}_0)=\pm \frac{\mathrm{2\Omega }}{\mathrm{\Omega }+{\mathrm{\Omega }}_0}\eta m_\mathrm{0z}\simeq \pm \eta m_\mathrm{0z},
\end{array}
\]

Therefore,

\begin{equation}\tag{S32}
g=\frac{1}{2}\mathrm{\Delta }{\mathrm{\Omega}_{\mathrm{gap}}}\gamma B_\text{s}=\frac{\gamma B_{\text{ex}}B_0}{2B_\text{s}+4B_{\text{ex}}}{\mathrm{cos} {\theta }_\text{B}}
\end{equation}
Note that this is only valid when ${\mathrm{\Delta }{\mathrm{\Omega }}_{\mathrm{gap}}\mathrm{\ll }\mathrm{\Omega }}_0$.

\section{Impact of the Mutual Spin Pumping Damping on the Coupling}
\begin{figure}[b]
 \includegraphics[]{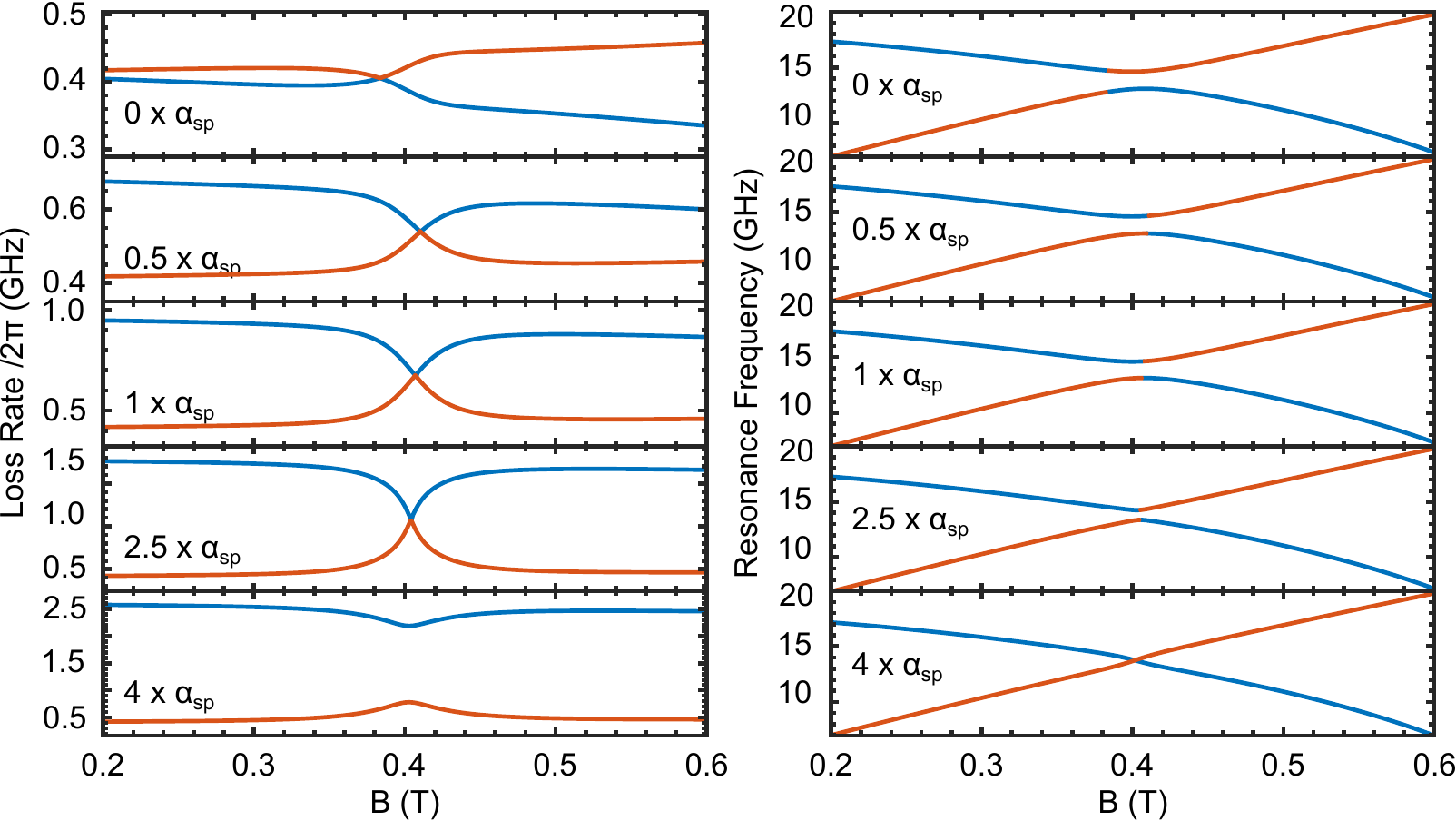}
 \caption{Simulated loss rate (left panel) and resonance frequency (right panel) of the acoustic (red) and optic (blue) modes as a function of applied magnetic field at an angle of $\theta_{\mathrm{B}} = 27\degree$ for different multiples of the mutual spin pumping damping $\alpha_{\mathrm{sp}}=0.0255$. These are produced by solving Eq.\,S22 with other parameters of $B_{\mathrm{S}} = 1.583\,\tesla$, $B_{\mathrm{ex},1} = 0.14\,\tesla$, $B_{\mathrm{ex,2}} = 0.0065\,\tesla$, $\gamma/2\pi = 29\,\giga\hertz/\tesla$ and $\alpha_0 = 0.0155$.  \label{S1}}
 \end{figure}
Our theory model allows to explore parameter regimes beyond experimental conditions. In an effort to understand the coupling of the two magnetic resonance modes in a SyAF, we investigated qualitatively the dependence of the damping parameters on the coupling strength. Especially, we focused on the effect of the damping arising from the mutual spin pumping between the two ferromagnets. This is because of our derived matrix form in Eq. S26 in which $\alpha_{\mathrm{sp}}$ exists in the off-diagonal term, strongly suggesting that this parameter can contribute to the coupling of optical and acoustic modes. Figure \ref{S1} shows the simulated loss rate and resonance frequency of the acoustic and optic modes as a function of
the magnetic field $B$ for $\theta_{\mathrm{B}}= 27\degree$. The plots show the results of the full eigenvalue problem defined in Eq.\,S22 for several values of $\alpha_{\mathrm{sp}}$. For zero $\alpha_{\mathrm{sp}}$, the loss rate of the acoustic and optic mode cross at a lower magnetic field than the point of minimal frequency separation between the resonance frequencies of the two modes. With increasing $\alpha_{\mathrm{sp}}$, the crossing of the loss rates shifts higher in magnetic field until it appears at the same field as the minimal frequency separation of the resonance frequencies. For the highest $\alpha_{\mathrm{sp}}$ the loss rate no longer cross and are separated. The dispersion of the resonance frequencies remains mostly unchanged for $\alpha_{\mathrm{sp}} = 0$ and $\alpha_{\mathrm{sp}} = 0.0255$. We observe that for higher $\alpha_{\mathrm{sp}}$, the coupling strength reduces and eventually goes to zero for the highest $\alpha_{\mathrm{sp}}$ we plot. From the change of the coupling strength with $\alpha_{\mathrm{sp}}$, we conclude that the coupling is partially mediated by spin currents. This change in the coupling behaviour can only be achieved by changing $\alpha_{\mathrm{sp}}$. In contrast, a change of the
Gilbert damping $\alpha_0$ has no influence on the coupling strength and only affects the loss rate of the modes. Our theory model suggests that the coupling between the acoustic and optic modes of a synthetic antiferromagnet is not fully described by a classical coupled harmonic oscillator model.

\end{document}